\documentclass[aps,prd]{revtex4}

\begin{document}
\title{Model Interacting Boson-Fermion Stars%
\footnote[3]{Report formerly published in the proceedings of XVI Brazilian National Meeting on Particles and Fields, Caxambu-MG, Brazil, October 24-28, 1995. This 2006 version corresponds to the original 1996 one, and has been specially prepared just to insertion in the ArXiv for register adjustment.}}

\author{Claudio M. G. de Sousa\footnote{At the time of publication of this congress report, on leave from Departamento de Fisica, Universidade de Brasilia, 70910-900, Brasilia-DF, Brazil, {\em and} International Centre of Condensed Matter Physics, Universidade de Brasilia, 70919-970, Brasilia-DF, Brazil. Present address:  Universidade Catolica de Brasilia, Diretoria de Fisica, 72022-900, Brasilia-DF, Brazil; electronic address: desousa@ucb.br} 
and J. L. Tomazelli\footnote{At the time of publication of this congress report, on leave from Instituto de Fisica Teorica-UNESP, Rua Pamplona 145, 01405-900, Sao Paulo-SP, Brazil. Present address: Dept. Fisica, CCFM, Universidade Federal de Santa Catarina, CP 476, CEP 88010-970, Santa Catarina, SC, Brazil; electronic address: tomazelli@fsc.ufsc.br} }
\affiliation{International Centre for Theoretical Physics, \\
               P.O.Box 586, 34100, Trieste, Italy}

\author{Vanda Silveira}
\affiliation{Departamento de Fisica, Universidade de Brasilia, 70910-900, Brasilia-DF, Brazil}

\date{Received March, 1996}

\maketitle

The more accepted cosmological models predict the existence of a hidden matter, the so called Dark Matter, without which the observed luminous mass density does not approximate the expected critical value. Astrophysical observations \cite{Kolb90} of gas clouds in spiral galaxies lead one to conclude that the observed luminous mass is not sufficient to explain its orbital velocity: the dark matter should be 3 to 10 times more abundant than the amount observed for the luminous one.

In spite of the fact they are known for a certain time \cite{Ruffini69}, compact equilibrium configurations of boson fields (boson stars) have recently arouse great interest due to the conjecture that the dark matter may be composed partially by bosonic particles, under the form of compact objects. Since its formation probably occurred in the early universe in the presence of fermions, one may expect that those structures also contain fermions.

This report gives an outline of current-current type interaction between%
self-gravitating bosons and fermions \cite{Sousa98}. Another approach for interaction in a boson-fermion star has been suggested \cite{Henriques89} in a different way. In this model one introduces contact interaction for bosons and fermions via
\begin{equation}
{\cal L}^{int} = \lambda J_{\mu} (\Phi ) j^{\mu} (\Psi ) 	    \label{01}
\end{equation}
where
\begin{equation}
J_{\mu} (\Phi ) = i \left( \Phi^* \partial_{\mu}\Phi -  \Phi \partial_{\mu}\Phi^* \right)
\end{equation}
\begin{equation}
j^{\mu} (\Psi ) = \bar\Psi \gamma^\mu \Psi
\end{equation}
which represent the boson and fermion currents, respectively. The current for the boson fields arises from  the usual Lagrangian for the boson stars
\begin{equation}
{\cal L} = \frac{R}{16\pi G} - \partial_{\mu}\Phi^* \partial^{\mu}\Phi - m^2 \Phi^* \Phi
\end{equation}
where
\begin{equation}
\Phi (r,\tau )=\phi (r) e^{-i\omega \tau}
\end{equation}
and $\omega$ is a frequency which determines the system energy. The cosmological time $\tau$ is given by the metric used, which is spherically symmetric chosen in the form
\begin{equation}
	ds^2 = -B(r)d \tau^2 + A(r) dr^2 +r^2 d\theta^2 + r^2 \sin^2 \theta d\varphi^2
\end{equation}

Fermions are introduced as a relativistic perfect fluid in accordance with the prescriptions for pure fermion and boson-fermion stars \cite{Henriques89}, \cite{Oppenheimer39}. The total energy-momentum tensor is given by the contribution of boson and fermion matter, and by the interaction term
\begin{equation}
	T_{\mu\nu}= T_{\mu\nu}^{B} + T_{\mu\nu}^{F} + T_{\mu\nu}^{int}
\end{equation}
where the labels $B$, $F$ and $`int´$ mean boson, fermion and interaction terms, respectively. Due to the metric only two components of the interaction energy-momentum tensor are used to determine the differential equations, namely $T_\tau^{\tau \,\,\, int}$ and $T_r^{r \,\,\, int}.$ If one considers $\phi (r)$ as real scalar field, as usually done for boson stars, $T_\tau^{\tau \,\,\, int}$ vanishes. This fact is not desirable since this component is representative of the changes in energy corresponding to the interaction. Hence, one is led to write the scalar field as a sum of real and imaginary parts $\phi (r)= \phi_1 (r) + i \phi_2 (r) $, with which%
$\left| \Phi \right|^2 = \left| \phi \right|^2 = \phi_1^2 + \phi_1^2 .$

The mixed configuration considered here takes boson field in its ground-state and the fermion fluid is allowed to display very slow radial velocities. This directly expresses that chemical potential for the Fermi gas is not to high in the kinetic energy scale, and so the Fermi surface allows only very slow velocities.

The interaction introduced by (\ref{01}) imposes many modifications in the configuration, one of those is a singularity in the evolution equations. Another interesting feature of this kind of interaction is that the ground-state energy of the configuration increases with $\lambda$ up to a certain value, where the energy starts decreasing while $\lambda$ increases. At this point of the system energy the real and imaginary parts of $\phi (r)$ display different behaviours, as shown in the figure.

The approach presented here can give rise to other kind of stellar objects, as stars of WIMP's \cite{Pisano96}, for which one takes into account interacting boson-fermion stars reinterpreted in the context of minimal supersymmetric standard model. In some sense an enhancement of the emission rate of gravitational waves can be expected due to the interaction within the star body and its neighbourhood.

\begin{figure}[htbp]
	\centering

\setlength{\unitlength}{0.240900pt}
\ifx\plotpoint\undefined\newsavebox{\plotpoint}\fi
\sbox{\plotpoint}{\rule[-0.200pt]{0.400pt}{0.400pt}}%
\begin{picture}(1500,900)(0,0)
\sbox{\plotpoint}{\rule[-0.200pt]{0.400pt}{0.400pt}}%
\put(60.0,40.0){\rule[-0.200pt]{332.201pt}{0.400pt}}
\put(1439.0,40.0){\rule[-0.200pt]{0.400pt}{197.538pt}}
\put(60.0,860.0){\rule[-0.200pt]{332.201pt}{0.400pt}}
\put(60.0,40.0){\rule[-0.200pt]{0.400pt}{197.538pt}}
\put(750,631){\makebox(0,0)[l]{$\phi_1 $}}
\put(750,356){\makebox(0,0)[l]{$\phi_2 $}}
\put(60,860){\usebox{\plotpoint}}
\put(74,858.67){\rule{3.373pt}{0.400pt}}
\multiput(74.00,859.17)(7.000,-1.000){2}{\rule{1.686pt}{0.400pt}}
\put(88,857.67){\rule{3.373pt}{0.400pt}}
\multiput(88.00,858.17)(7.000,-1.000){2}{\rule{1.686pt}{0.400pt}}
\put(102,856.17){\rule{2.900pt}{0.400pt}}
\multiput(102.00,857.17)(7.981,-2.000){2}{\rule{1.450pt}{0.400pt}}
\put(116,854.17){\rule{2.900pt}{0.400pt}}
\multiput(116.00,855.17)(7.981,-2.000){2}{\rule{1.450pt}{0.400pt}}
\put(130,852.17){\rule{2.900pt}{0.400pt}}
\multiput(130.00,853.17)(7.981,-2.000){2}{\rule{1.450pt}{0.400pt}}
\multiput(144.00,850.95)(2.918,-0.447){3}{\rule{1.967pt}{0.108pt}}
\multiput(144.00,851.17)(9.918,-3.000){2}{\rule{0.983pt}{0.400pt}}
\multiput(158.00,847.95)(2.695,-0.447){3}{\rule{1.833pt}{0.108pt}}
\multiput(158.00,848.17)(9.195,-3.000){2}{\rule{0.917pt}{0.400pt}}
\multiput(171.00,844.95)(2.918,-0.447){3}{\rule{1.967pt}{0.108pt}}
\multiput(171.00,845.17)(9.918,-3.000){2}{\rule{0.983pt}{0.400pt}}
\multiput(185.00,841.94)(1.943,-0.468){5}{\rule{1.500pt}{0.113pt}}
\multiput(185.00,842.17)(10.887,-4.000){2}{\rule{0.750pt}{0.400pt}}
\multiput(199.00,837.94)(1.943,-0.468){5}{\rule{1.500pt}{0.113pt}}
\multiput(199.00,838.17)(10.887,-4.000){2}{\rule{0.750pt}{0.400pt}}
\multiput(213.00,833.93)(1.489,-0.477){7}{\rule{1.220pt}{0.115pt}}
\multiput(213.00,834.17)(11.468,-5.000){2}{\rule{0.610pt}{0.400pt}}
\multiput(227.00,828.93)(1.489,-0.477){7}{\rule{1.220pt}{0.115pt}}
\multiput(227.00,829.17)(11.468,-5.000){2}{\rule{0.610pt}{0.400pt}}
\multiput(241.00,823.93)(1.489,-0.477){7}{\rule{1.220pt}{0.115pt}}
\multiput(241.00,824.17)(11.468,-5.000){2}{\rule{0.610pt}{0.400pt}}
\multiput(255.00,818.93)(1.214,-0.482){9}{\rule{1.033pt}{0.116pt}}
\multiput(255.00,819.17)(11.855,-6.000){2}{\rule{0.517pt}{0.400pt}}
\multiput(269.00,812.93)(1.214,-0.482){9}{\rule{1.033pt}{0.116pt}}
\multiput(269.00,813.17)(11.855,-6.000){2}{\rule{0.517pt}{0.400pt}}
\multiput(283.00,806.93)(1.214,-0.482){9}{\rule{1.033pt}{0.116pt}}
\multiput(283.00,807.17)(11.855,-6.000){2}{\rule{0.517pt}{0.400pt}}
\multiput(297.00,800.93)(1.026,-0.485){11}{\rule{0.900pt}{0.117pt}}
\multiput(297.00,801.17)(12.132,-7.000){2}{\rule{0.450pt}{0.400pt}}
\multiput(311.00,793.93)(1.026,-0.485){11}{\rule{0.900pt}{0.117pt}}
\multiput(311.00,794.17)(12.132,-7.000){2}{\rule{0.450pt}{0.400pt}}
\multiput(325.00,786.93)(1.026,-0.485){11}{\rule{0.900pt}{0.117pt}}
\multiput(325.00,787.17)(12.132,-7.000){2}{\rule{0.450pt}{0.400pt}}
\multiput(339.00,779.93)(0.890,-0.488){13}{\rule{0.800pt}{0.117pt}}
\multiput(339.00,780.17)(12.340,-8.000){2}{\rule{0.400pt}{0.400pt}}
\multiput(353.00,771.93)(0.824,-0.488){13}{\rule{0.750pt}{0.117pt}}
\multiput(353.00,772.17)(11.443,-8.000){2}{\rule{0.375pt}{0.400pt}}
\multiput(366.00,763.93)(0.890,-0.488){13}{\rule{0.800pt}{0.117pt}}
\multiput(366.00,764.17)(12.340,-8.000){2}{\rule{0.400pt}{0.400pt}}
\multiput(380.00,755.93)(0.890,-0.488){13}{\rule{0.800pt}{0.117pt}}
\multiput(380.00,756.17)(12.340,-8.000){2}{\rule{0.400pt}{0.400pt}}
\multiput(394.00,747.93)(0.786,-0.489){15}{\rule{0.722pt}{0.118pt}}
\multiput(394.00,748.17)(12.501,-9.000){2}{\rule{0.361pt}{0.400pt}}
\multiput(408.00,738.93)(0.786,-0.489){15}{\rule{0.722pt}{0.118pt}}
\multiput(408.00,739.17)(12.501,-9.000){2}{\rule{0.361pt}{0.400pt}}
\multiput(422.00,729.93)(0.786,-0.489){15}{\rule{0.722pt}{0.118pt}}
\multiput(422.00,730.17)(12.501,-9.000){2}{\rule{0.361pt}{0.400pt}}
\multiput(436.00,720.92)(0.704,-0.491){17}{\rule{0.660pt}{0.118pt}}
\multiput(436.00,721.17)(12.630,-10.000){2}{\rule{0.330pt}{0.400pt}}
\multiput(450.00,710.93)(0.786,-0.489){15}{\rule{0.722pt}{0.118pt}}
\multiput(450.00,711.17)(12.501,-9.000){2}{\rule{0.361pt}{0.400pt}}
\multiput(464.00,701.92)(0.704,-0.491){17}{\rule{0.660pt}{0.118pt}}
\multiput(464.00,702.17)(12.630,-10.000){2}{\rule{0.330pt}{0.400pt}}
\multiput(478.00,691.92)(0.704,-0.491){17}{\rule{0.660pt}{0.118pt}}
\multiput(478.00,692.17)(12.630,-10.000){2}{\rule{0.330pt}{0.400pt}}
\multiput(492.00,681.92)(0.704,-0.491){17}{\rule{0.660pt}{0.118pt}}
\multiput(492.00,682.17)(12.630,-10.000){2}{\rule{0.330pt}{0.400pt}}
\multiput(506.00,671.92)(0.637,-0.492){19}{\rule{0.609pt}{0.118pt}}
\multiput(506.00,672.17)(12.736,-11.000){2}{\rule{0.305pt}{0.400pt}}
\multiput(520.00,660.92)(0.637,-0.492){19}{\rule{0.609pt}{0.118pt}}
\multiput(520.00,661.17)(12.736,-11.000){2}{\rule{0.305pt}{0.400pt}}
\multiput(534.00,649.92)(0.704,-0.491){17}{\rule{0.660pt}{0.118pt}}
\multiput(534.00,650.17)(12.630,-10.000){2}{\rule{0.330pt}{0.400pt}}
\multiput(548.00,639.92)(0.590,-0.492){19}{\rule{0.573pt}{0.118pt}}
\multiput(548.00,640.17)(11.811,-11.000){2}{\rule{0.286pt}{0.400pt}}
\multiput(561.00,628.92)(0.637,-0.492){19}{\rule{0.609pt}{0.118pt}}
\multiput(561.00,629.17)(12.736,-11.000){2}{\rule{0.305pt}{0.400pt}}
\multiput(575.00,617.92)(0.637,-0.492){19}{\rule{0.609pt}{0.118pt}}
\multiput(575.00,618.17)(12.736,-11.000){2}{\rule{0.305pt}{0.400pt}}
\multiput(589.00,606.92)(0.582,-0.492){21}{\rule{0.567pt}{0.119pt}}
\multiput(589.00,607.17)(12.824,-12.000){2}{\rule{0.283pt}{0.400pt}}
\multiput(603.00,594.92)(0.637,-0.492){19}{\rule{0.609pt}{0.118pt}}
\multiput(603.00,595.17)(12.736,-11.000){2}{\rule{0.305pt}{0.400pt}}
\multiput(617.00,583.92)(0.637,-0.492){19}{\rule{0.609pt}{0.118pt}}
\multiput(617.00,584.17)(12.736,-11.000){2}{\rule{0.305pt}{0.400pt}}
\multiput(631.00,572.92)(0.582,-0.492){21}{\rule{0.567pt}{0.119pt}}
\multiput(631.00,573.17)(12.824,-12.000){2}{\rule{0.283pt}{0.400pt}}
\multiput(645.00,560.92)(0.637,-0.492){19}{\rule{0.609pt}{0.118pt}}
\multiput(645.00,561.17)(12.736,-11.000){2}{\rule{0.305pt}{0.400pt}}
\multiput(659.00,549.92)(0.582,-0.492){21}{\rule{0.567pt}{0.119pt}}
\multiput(659.00,550.17)(12.824,-12.000){2}{\rule{0.283pt}{0.400pt}}
\multiput(673.00,537.92)(0.582,-0.492){21}{\rule{0.567pt}{0.119pt}}
\multiput(673.00,538.17)(12.824,-12.000){2}{\rule{0.283pt}{0.400pt}}
\multiput(687.00,525.92)(0.637,-0.492){19}{\rule{0.609pt}{0.118pt}}
\multiput(687.00,526.17)(12.736,-11.000){2}{\rule{0.305pt}{0.400pt}}
\multiput(701.00,514.92)(0.582,-0.492){21}{\rule{0.567pt}{0.119pt}}
\multiput(701.00,515.17)(12.824,-12.000){2}{\rule{0.283pt}{0.400pt}}
\multiput(715.00,502.92)(0.582,-0.492){21}{\rule{0.567pt}{0.119pt}}
\multiput(715.00,503.17)(12.824,-12.000){2}{\rule{0.283pt}{0.400pt}}
\multiput(729.00,490.92)(0.637,-0.492){19}{\rule{0.609pt}{0.118pt}}
\multiput(729.00,491.17)(12.736,-11.000){2}{\rule{0.305pt}{0.400pt}}
\multiput(743.00,479.92)(0.539,-0.492){21}{\rule{0.533pt}{0.119pt}}
\multiput(743.00,480.17)(11.893,-12.000){2}{\rule{0.267pt}{0.400pt}}
\multiput(756.00,467.92)(0.582,-0.492){21}{\rule{0.567pt}{0.119pt}}
\multiput(756.00,468.17)(12.824,-12.000){2}{\rule{0.283pt}{0.400pt}}
\multiput(770.00,455.92)(0.582,-0.492){21}{\rule{0.567pt}{0.119pt}}
\multiput(770.00,456.17)(12.824,-12.000){2}{\rule{0.283pt}{0.400pt}}
\multiput(784.00,443.92)(0.637,-0.492){19}{\rule{0.609pt}{0.118pt}}
\multiput(784.00,444.17)(12.736,-11.000){2}{\rule{0.305pt}{0.400pt}}
\multiput(798.00,432.92)(0.582,-0.492){21}{\rule{0.567pt}{0.119pt}}
\multiput(798.00,433.17)(12.824,-12.000){2}{\rule{0.283pt}{0.400pt}}
\multiput(812.00,420.92)(0.637,-0.492){19}{\rule{0.609pt}{0.118pt}}
\multiput(812.00,421.17)(12.736,-11.000){2}{\rule{0.305pt}{0.400pt}}
\multiput(826.00,409.92)(0.582,-0.492){21}{\rule{0.567pt}{0.119pt}}
\multiput(826.00,410.17)(12.824,-12.000){2}{\rule{0.283pt}{0.400pt}}
\multiput(840.00,397.92)(0.637,-0.492){19}{\rule{0.609pt}{0.118pt}}
\multiput(840.00,398.17)(12.736,-11.000){2}{\rule{0.305pt}{0.400pt}}
\multiput(854.00,386.92)(0.637,-0.492){19}{\rule{0.609pt}{0.118pt}}
\multiput(854.00,387.17)(12.736,-11.000){2}{\rule{0.305pt}{0.400pt}}
\multiput(868.00,375.92)(0.582,-0.492){21}{\rule{0.567pt}{0.119pt}}
\multiput(868.00,376.17)(12.824,-12.000){2}{\rule{0.283pt}{0.400pt}}
\multiput(882.00,363.92)(0.637,-0.492){19}{\rule{0.609pt}{0.118pt}}
\multiput(882.00,364.17)(12.736,-11.000){2}{\rule{0.305pt}{0.400pt}}
\multiput(896.00,352.92)(0.637,-0.492){19}{\rule{0.609pt}{0.118pt}}
\multiput(896.00,353.17)(12.736,-11.000){2}{\rule{0.305pt}{0.400pt}}
\multiput(910.00,341.92)(0.704,-0.491){17}{\rule{0.660pt}{0.118pt}}
\multiput(910.00,342.17)(12.630,-10.000){2}{\rule{0.330pt}{0.400pt}}
\multiput(924.00,331.92)(0.637,-0.492){19}{\rule{0.609pt}{0.118pt}}
\multiput(924.00,332.17)(12.736,-11.000){2}{\rule{0.305pt}{0.400pt}}
\multiput(938.00,320.92)(0.590,-0.492){19}{\rule{0.573pt}{0.118pt}}
\multiput(938.00,321.17)(11.811,-11.000){2}{\rule{0.286pt}{0.400pt}}
\multiput(951.00,309.92)(0.704,-0.491){17}{\rule{0.660pt}{0.118pt}}
\multiput(951.00,310.17)(12.630,-10.000){2}{\rule{0.330pt}{0.400pt}}
\multiput(965.00,299.92)(0.637,-0.492){19}{\rule{0.609pt}{0.118pt}}
\multiput(965.00,300.17)(12.736,-11.000){2}{\rule{0.305pt}{0.400pt}}
\multiput(979.00,288.92)(0.704,-0.491){17}{\rule{0.660pt}{0.118pt}}
\multiput(979.00,289.17)(12.630,-10.000){2}{\rule{0.330pt}{0.400pt}}
\multiput(993.00,278.92)(0.704,-0.491){17}{\rule{0.660pt}{0.118pt}}
\multiput(993.00,279.17)(12.630,-10.000){2}{\rule{0.330pt}{0.400pt}}
\multiput(1007.00,268.92)(0.704,-0.491){17}{\rule{0.660pt}{0.118pt}}
\multiput(1007.00,269.17)(12.630,-10.000){2}{\rule{0.330pt}{0.400pt}}
\multiput(1021.00,258.92)(0.704,-0.491){17}{\rule{0.660pt}{0.118pt}}
\multiput(1021.00,259.17)(12.630,-10.000){2}{\rule{0.330pt}{0.400pt}}
\multiput(1035.00,248.92)(0.704,-0.491){17}{\rule{0.660pt}{0.118pt}}
\multiput(1035.00,249.17)(12.630,-10.000){2}{\rule{0.330pt}{0.400pt}}
\multiput(1049.00,238.93)(0.786,-0.489){15}{\rule{0.722pt}{0.118pt}}
\multiput(1049.00,239.17)(12.501,-9.000){2}{\rule{0.361pt}{0.400pt}}
\multiput(1063.00,229.93)(0.786,-0.489){15}{\rule{0.722pt}{0.118pt}}
\multiput(1063.00,230.17)(12.501,-9.000){2}{\rule{0.361pt}{0.400pt}}
\multiput(1077.00,220.92)(0.704,-0.491){17}{\rule{0.660pt}{0.118pt}}
\multiput(1077.00,221.17)(12.630,-10.000){2}{\rule{0.330pt}{0.400pt}}
\multiput(1091.00,210.93)(0.786,-0.489){15}{\rule{0.722pt}{0.118pt}}
\multiput(1091.00,211.17)(12.501,-9.000){2}{\rule{0.361pt}{0.400pt}}
\multiput(1105.00,201.93)(0.890,-0.488){13}{\rule{0.800pt}{0.117pt}}
\multiput(1105.00,202.17)(12.340,-8.000){2}{\rule{0.400pt}{0.400pt}}
\multiput(1119.00,193.93)(0.786,-0.489){15}{\rule{0.722pt}{0.118pt}}
\multiput(1119.00,194.17)(12.501,-9.000){2}{\rule{0.361pt}{0.400pt}}
\multiput(1133.00,184.93)(0.824,-0.488){13}{\rule{0.750pt}{0.117pt}}
\multiput(1133.00,185.17)(11.443,-8.000){2}{\rule{0.375pt}{0.400pt}}
\multiput(1146.00,176.93)(0.786,-0.489){15}{\rule{0.722pt}{0.118pt}}
\multiput(1146.00,177.17)(12.501,-9.000){2}{\rule{0.361pt}{0.400pt}}
\multiput(1160.00,167.93)(0.890,-0.488){13}{\rule{0.800pt}{0.117pt}}
\multiput(1160.00,168.17)(12.340,-8.000){2}{\rule{0.400pt}{0.400pt}}
\multiput(1174.00,159.93)(0.890,-0.488){13}{\rule{0.800pt}{0.117pt}}
\multiput(1174.00,160.17)(12.340,-8.000){2}{\rule{0.400pt}{0.400pt}}
\multiput(1188.00,151.93)(0.890,-0.488){13}{\rule{0.800pt}{0.117pt}}
\multiput(1188.00,152.17)(12.340,-8.000){2}{\rule{0.400pt}{0.400pt}}
\multiput(1202.00,143.93)(1.026,-0.485){11}{\rule{0.900pt}{0.117pt}}
\multiput(1202.00,144.17)(12.132,-7.000){2}{\rule{0.450pt}{0.400pt}}
\multiput(1216.00,136.93)(0.890,-0.488){13}{\rule{0.800pt}{0.117pt}}
\multiput(1216.00,137.17)(12.340,-8.000){2}{\rule{0.400pt}{0.400pt}}
\multiput(1230.00,128.93)(1.026,-0.485){11}{\rule{0.900pt}{0.117pt}}
\multiput(1230.00,129.17)(12.132,-7.000){2}{\rule{0.450pt}{0.400pt}}
\multiput(1244.00,121.93)(1.026,-0.485){11}{\rule{0.900pt}{0.117pt}}
\multiput(1244.00,122.17)(12.132,-7.000){2}{\rule{0.450pt}{0.400pt}}
\multiput(1258.00,114.93)(1.026,-0.485){11}{\rule{0.900pt}{0.117pt}}
\multiput(1258.00,115.17)(12.132,-7.000){2}{\rule{0.450pt}{0.400pt}}
\multiput(1272.00,107.93)(1.026,-0.485){11}{\rule{0.900pt}{0.117pt}}
\multiput(1272.00,108.17)(12.132,-7.000){2}{\rule{0.450pt}{0.400pt}}
\multiput(1286.00,100.93)(1.214,-0.482){9}{\rule{1.033pt}{0.116pt}}
\multiput(1286.00,101.17)(11.855,-6.000){2}{\rule{0.517pt}{0.400pt}}
\multiput(1300.00,94.93)(1.214,-0.482){9}{\rule{1.033pt}{0.116pt}}
\multiput(1300.00,95.17)(11.855,-6.000){2}{\rule{0.517pt}{0.400pt}}
\multiput(1314.00,88.93)(1.026,-0.485){11}{\rule{0.900pt}{0.117pt}}
\multiput(1314.00,89.17)(12.132,-7.000){2}{\rule{0.450pt}{0.400pt}}
\multiput(1328.00,81.93)(1.123,-0.482){9}{\rule{0.967pt}{0.116pt}}
\multiput(1328.00,82.17)(10.994,-6.000){2}{\rule{0.483pt}{0.400pt}}
\multiput(1341.00,75.93)(1.489,-0.477){7}{\rule{1.220pt}{0.115pt}}
\multiput(1341.00,76.17)(11.468,-5.000){2}{\rule{0.610pt}{0.400pt}}
\multiput(1355.00,70.93)(1.214,-0.482){9}{\rule{1.033pt}{0.116pt}}
\multiput(1355.00,71.17)(11.855,-6.000){2}{\rule{0.517pt}{0.400pt}}
\multiput(1369.00,64.93)(1.214,-0.482){9}{\rule{1.033pt}{0.116pt}}
\multiput(1369.00,65.17)(11.855,-6.000){2}{\rule{0.517pt}{0.400pt}}
\multiput(1383.00,58.93)(1.489,-0.477){7}{\rule{1.220pt}{0.115pt}}
\multiput(1383.00,59.17)(11.468,-5.000){2}{\rule{0.610pt}{0.400pt}}
\multiput(1397.00,53.93)(1.489,-0.477){7}{\rule{1.220pt}{0.115pt}}
\multiput(1397.00,54.17)(11.468,-5.000){2}{\rule{0.610pt}{0.400pt}}
\multiput(1411.00,48.93)(1.489,-0.477){7}{\rule{1.220pt}{0.115pt}}
\multiput(1411.00,49.17)(11.468,-5.000){2}{\rule{0.610pt}{0.400pt}}
\multiput(1425.00,43.93)(1.489,-0.477){7}{\rule{1.220pt}{0.115pt}}
\multiput(1425.00,44.17)(11.468,-5.000){2}{\rule{0.610pt}{0.400pt}}
\put(60.0,860.0){\rule[-0.200pt]{3.373pt}{0.400pt}}
\put(60,860){\usebox{\plotpoint}}
\put(60.00,860.00){\usebox{\plotpoint}}
\put(80.74,859.52){\usebox{\plotpoint}}
\put(101.44,858.04){\usebox{\plotpoint}}
\put(122.13,857.12){\usebox{\plotpoint}}
\put(142.78,855.09){\usebox{\plotpoint}}
\put(163.32,852.18){\usebox{\plotpoint}}
\put(183.86,849.16){\usebox{\plotpoint}}
\put(204.34,845.86){\usebox{\plotpoint}}
\put(224.63,841.51){\usebox{\plotpoint}}
\put(244.93,837.16){\usebox{\plotpoint}}
\put(265.05,832.13){\usebox{\plotpoint}}
\put(285.24,827.36){\usebox{\plotpoint}}
\put(305.03,821.13){\usebox{\plotpoint}}
\put(324.86,815.04){\usebox{\plotpoint}}
\put(344.70,808.96){\usebox{\plotpoint}}
\put(364.15,801.71){\usebox{\plotpoint}}
\put(383.59,794.46){\usebox{\plotpoint}}
\put(402.88,786.83){\usebox{\plotpoint}}
\put(422.08,778.97){\usebox{\plotpoint}}
\put(441.16,770.79){\usebox{\plotpoint}}
\put(460.23,762.61){\usebox{\plotpoint}}
\put(478.92,753.60){\usebox{\plotpoint}}
\put(497.84,745.08){\usebox{\plotpoint}}
\put(516.40,735.80){\usebox{\plotpoint}}
\put(534.97,726.52){\usebox{\plotpoint}}
\put(553.27,716.76){\usebox{\plotpoint}}
\put(571.44,706.78){\usebox{\plotpoint}}
\put(589.59,696.71){\usebox{\plotpoint}}
\put(608.00,687.14){\usebox{\plotpoint}}
\put(626.02,676.85){\usebox{\plotpoint}}
\put(644.04,666.55){\usebox{\plotpoint}}
\put(661.97,656.09){\usebox{\plotpoint}}
\put(679.63,645.21){\usebox{\plotpoint}}
\put(697.65,634.91){\usebox{\plotpoint}}
\put(715.22,623.86){\usebox{\plotpoint}}
\put(732.79,612.83){\usebox{\plotpoint}}
\put(750.40,601.88){\usebox{\plotpoint}}
\put(767.73,590.46){\usebox{\plotpoint}}
\put(785.23,579.30){\usebox{\plotpoint}}
\put(803.08,568.73){\usebox{\plotpoint}}
\put(820.54,557.51){\usebox{\plotpoint}}
\put(838.00,546.28){\usebox{\plotpoint}}
\put(855.46,535.06){\usebox{\plotpoint}}
\put(872.76,523.60){\usebox{\plotpoint}}
\put(889.91,511.92){\usebox{\plotpoint}}
\put(907.37,500.69){\usebox{\plotpoint}}
\put(924.83,489.47){\usebox{\plotpoint}}
\put(942.19,478.10){\usebox{\plotpoint}}
\put(959.44,466.57){\usebox{\plotpoint}}
\put(976.90,455.35){\usebox{\plotpoint}}
\put(994.36,444.12){\usebox{\plotpoint}}
\put(1011.82,432.90){\usebox{\plotpoint}}
\put(1029.28,421.68){\usebox{\plotpoint}}
\put(1046.74,410.45){\usebox{\plotpoint}}
\put(1064.20,399.23){\usebox{\plotpoint}}
\put(1081.66,388.01){\usebox{\plotpoint}}
\put(1099.38,377.21){\usebox{\plotpoint}}
\put(1117.01,366.28){\usebox{\plotpoint}}
\put(1134.49,355.08){\usebox{\plotpoint}}
\put(1152.09,344.09){\usebox{\plotpoint}}
\put(1169.86,333.37){\usebox{\plotpoint}}
\put(1187.44,322.36){\usebox{\plotpoint}}
\put(1205.45,312.03){\usebox{\plotpoint}}
\put(1223.47,301.73){\usebox{\plotpoint}}
\put(1241.49,291.43){\usebox{\plotpoint}}
\put(1259.51,281.14){\usebox{\plotpoint}}
\put(1277.53,270.84){\usebox{\plotpoint}}
\put(1295.55,260.54){\usebox{\plotpoint}}
\put(1313.98,251.01){\usebox{\plotpoint}}
\put(1332.06,240.81){\usebox{\plotpoint}}
\put(1350.20,230.74){\usebox{\plotpoint}}
\put(1368.62,221.19){\usebox{\plotpoint}}
\put(1387.19,211.91){\usebox{\plotpoint}}
\put(1405.75,202.62){\usebox{\plotpoint}}
\put(1424.32,193.34){\usebox{\plotpoint}}
\put(1439,186){\usebox{\plotpoint}}
\put(60.0,40.0){\rule[-0.200pt]{332.201pt}{0.400pt}}
\put(1439.0,40.0){\rule[-0.200pt]{0.400pt}{197.538pt}}
\put(60.0,860.0){\rule[-0.200pt]{332.201pt}{0.400pt}}
\put(60.0,40.0){\rule[-0.200pt]{0.400pt}{197.538pt}}
\end{picture}

	\caption{Plotting showing the splitting between $\phi_1$ and $\phi_2$, respectively the real and imaginary parts of the scalar field $\phi (r)$}
	\label{fig1}
\end{figure}
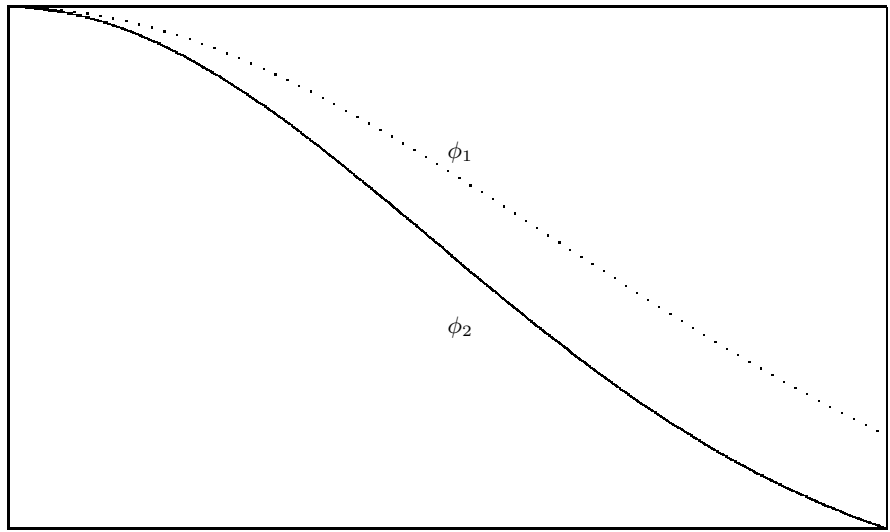

\end{document}